\documentclass[pra,aps,showpacs,superscriptaddress,twocolumn]{revtex4}
\usepackage{amsmath,amssymb}
\usepackage{graphicx}

\begin{document}

\title{Polarisation rotation of slow light with orbital angular momentum in ultracold atomic gases}

\date{\today{}}

\author{Julius Ruseckas}
\affiliation{Institute of Theoretical Physics and Astronomy of Vilnius
University, A.\ Go\v{s}tauto 12, 01108 Vilnius, Lithuania}

\author{Gediminas Juzeli\=unas}
\affiliation{Institute of Theoretical Physics and Astronomy of Vilnius
University, A.\ Go\v{s}tauto 12, 01108 Vilnius, Lithuania}

\author{Patrik \"Ohberg}
\affiliation{SUPA, Department of Physics, Heriot-Watt University, Edinburgh EH14 4AS, UK}

\author{Stephen M. Barnett }
\affiliation{SUPA, Department of Physics, University of Strathclyde, Glasgow G4 0NG, UK}

\begin{abstract}
We consider the propagation of slow light with an orbital angular momentum
(OAM) in a moving atomic medium. We have derived a general equation of motion
and applied it in analysing propagation of slow light with an OAM in a rotating
medium, such as a vortex lattice.  We have shown that the OAM of slow light
manifests itself in a rotation of the polarisation plane of linearly polarised
light. To extract a pure rotational phase shift, we suggest to measure a
difference in the angle of the polarisation plane rotation by two consecutive
light beams with opposite OAM. The differential angle $\Delta\alpha_{\ell}$ is
proportional to the rotation frequency of the medium $\omega_{\mathrm{rot}}$
and the winding number $\ell$ of light, and is inversely proportional to the
group velocity of light. For slow light the angle $\Delta\alpha_{\ell}$ should
be large enough to be detectable. The effect can be used as a tool for
measuring the rotation frequency $\omega_{\mathrm{rot}}$ of the medium.
\end{abstract}


\maketitle

\section{Introduction}

Light can be slowed down by seven orders of magnitude using the
Electromagnetically Induced Transparency (EIT) \cite{hau99,kash99,budker99}.
The EIT makes a resonant, opaque medium transparent by means of quantum
interference between the optical transitions induced by the control and probe
laser fields. As a result, a weak probe pulse travels slowly and almost without
a dissipation in a resonant medium controlled by another laser
\cite{harris90,boller91,arimondo96:_progr_optic,harris97,scully97:_quant_optic,bergmann98,hau99,kash99,budker99}.
Electromagnetically induced transparency was shown not only to slow down
dramatically  laser pulses \cite{hau99,kash99,budker99} but also to store them
\cite{liu01,phillips01} in atomic gases. Following the proposal of
Ref.~\cite{fleischhauer00}, storage and release of a probe pulse has been
demonstrated \cite{liu01,phillips01,turukhin01} by dynamically changing the
intensity of the control laser. The possibility to coherently control the
propagation of quantum light pulses opens up interesting applications such as
generation of nonclassical states in atomic ensembles and reversible quantum
memories for light waves
\cite{lukin00,juzeliunas02,lukin03,fleischhauer05,ginsberg07}.

Associated with EIT is a dramatic modification of the reflective properties of
the medium. In this paper we analyse the manifestation of the orbital angular
momentum (OAM) of light in such a highly dispersive medium.  During the last
decade optical OAM has received a great deal of interest
\cite{allen99,allen03}. However most studies on the topic deal with ordinary
(not slow) light. Here we consider the propagation of a slow light with an OAM
in a moving atomic medium, the emphasis being placed on a rotating medium, such
as a vortex lattice in an atomic Bose-Einstein condensate (BEC).  Unlike the
previous studies on the slow light in moving media
\cite{leonhardt00,fiurasek02,oehberg02,fleischhauer02-2,juzeliunas03,carusotto03,zimmer04},
we allow the light to carry an OAM along the propagation direction. We show
that the OAM of slow light manifests itself in a rotation of the polarisation
plane of a linearly polarised light. To extract a pure rotational phase shift,
we suggest to measure a difference in the angles of the  rotation of the
polarisation plane for two light beams characterised by opposite OAM.  The
differential angle $\Delta\alpha_\ell$ is proportional to the rotation
frequency of the medium $\omega_{\mathrm{rot}}$, as well as to the OAM of
light.  The effect can be used as a tool for measuring the rotation frequency
$\omega_{\mathrm{rot}}$ of the medium and/or the OAM of slow light.  We should
note that rotational phase shifts have been shown to lead to rotation of an
image by propagation through a rotating medium (\cite{padgett06} and references
therein).  This suggests that such effects might also be observable in the
configuration discussed here.

\begin{figure}[ht]
\includegraphics[width=8.5cm]{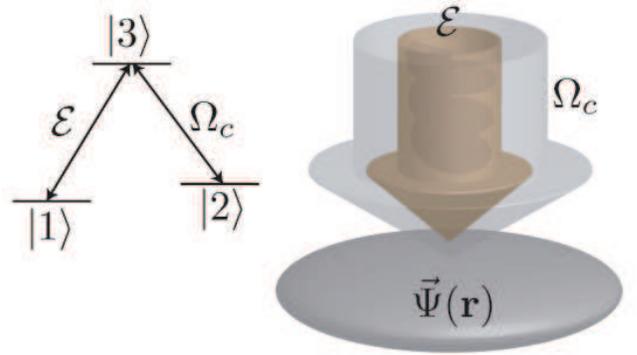}
\caption{Slow light propagating in a medium of $\Lambda$-type atoms.}
\label{fig1}
\end{figure}

\section{Formulation}

Consider an ensemble of $\Lambda$-type atoms characterized by two hyper-fine
ground levels $1$ and $2$, as well as an electronic excited level $3$ (Fig.~1).
The translational motion of atoms is described by the three component or vector
operator for the translational motion, $\vec \Psi(\mathbf{r})$, in which the
three components, $\Psi_1\equiv\Psi_1(\mathbf{r},t)$,
$\Psi_2\equiv\Psi_2(\mathbf{r},t)$ and $\Psi_3\equiv\Psi_3(\mathbf{r},t)$,
correspond to the three internal atomic states.  The quantum nature of the
atoms comprising the medium will, of course, determine whether these field
operators obey Bose-Einstein or Fermi-Dirac commutation relations.  The atoms
interact with two light fields: a strong classical control laser induces a
transition $|2\rangle\rightarrow|3\rangle$ and a weaker quantum probe field,
which drives a transition $|1\rangle\rightarrow|3\rangle$, see Fig.~1.

\subsection{Equation for the probe field}

The electric field of the probe beam is
\begin{equation}
\mathbf{E}(\mathbf{r},t)=\hat{\mathbf{e}}\sqrt{\frac{\hbar\omega}{
2\varepsilon_0}}\mathcal{E}(\mathbf{r},t)e^{-i\omega t}+\mathrm{H.c.}\:,
\label{eq:E}
\end{equation}
where $\omega=ck$ is the central frequency for probe photons,
$\mathbf{k}=\hat{\mathbf{z}}k$ is the wave-vector and
$\hat{\mathbf{e}}\perp\hat{\mathbf{z}}$ is the unit polarization vector. This
field may be considered as either a classical amplitude or a quantum operator.
We have chosen the dimensions of the electric field amplitude $\mathcal{E}$ so
that its squared modulus represents the number density of probe photons.

The probe field $\mathbf{E}(\mathbf{r},t)$ obeys the following wave-equation
\begin{equation}
c^2\nabla^2\mathbf{E}-\frac{\partial^2\mathbf{E}}{\partial
t^2}=\frac{1}{\varepsilon_0}\frac{\partial^2\mathbf{P}}{\partial
t^2}\:,
\label{eq:wave-equation}
\end{equation}
where
\begin{equation}
\mathbf{P}=\hat{\mathbf{e}}\mu\Psi_1^{\dagger}\Psi_3+\mathrm{H.c.}
\label{eq:P}
\end{equation}
is the polarisation field of atoms, $\mu$ being the dipole moment for the
atomic transition $|1\rangle\rightarrow|3\rangle$.

Let us introduce the slowly varying matter field-operators:
\begin{eqnarray}
\Phi_1&=&\Psi_1e^{i\omega_1t}\\
\Phi_2&=&\Psi_2e^{i(\omega_1+\omega-\omega_c)t}\\
\Phi_3&=&\Psi_3e^{i(\omega_1+\omega)t} 
\end{eqnarray}
where $\hbar\omega_1$ is the energy of the atomic ground state $1$ and
$\omega_c$ is the frequency of the control field. The probe field is
quasi-monochromatic, so that its amplitude
$\mathcal{E}\equiv\mathcal{E}(\mathbf{r},t)$ varies very slowly in time during
an optical cycle. In this case the following equation holds for the
slowly-varying amplitude of the probe field:
\begin{equation}
\left(\frac{\partial}{\partial
t}-i\frac{c^2}{2\omega}\nabla^2-i\frac{\omega}{2}\right)\mathcal{E}=ig\Phi_1^{
\dag}\Phi_3\:,
\label{eq:E-equation}
\end{equation}
where the parameter
\begin{equation}
g=\mu\sqrt{\omega/2\varepsilon_0\hbar}
\label{eq:g}
\end{equation}
characterizes the strength of coupling of the probe field with the atoms. Note
that, unlike in the usual treatment of slow light
\cite{lukin00,juzeliunas02,lukin03,fleischhauer05,fleischhauer02-2,juzeliunas03,carusotto03,zimmer04},
we have retained the second-order derivative $\nabla^2$ in the equation of
motion (\ref{eq:E-equation}). This allows us to account for the fast changes of
$\mathcal{E}$ in a direction perpendicular to the wave-vector $\mathbf{k}$,
i.e. in the $xy$-plane. Therefore, our analysis can be applied to the twisted
beams of light $\mathcal{E}(\mathbf{r},t)\sim\exp(i\ell \varphi)$ carrying an
OAM of $\hbar \ell$ per photon.

\subsection{Equations for the matter field-operators}

In the following we shall adapt a semiclassical picture in which both
electromagnetic and matter field operators are replaced by c-numbers. The
equations for the matter fields are then:
\begin{eqnarray}
\hat{K}\Phi_1 & = & V_1(\mathbf{r})\Phi_1-\hbar
g\mathcal{E}^{*}\Phi_3,
\label{eq:phi1}
\\\hat{K}\Phi_3 & = &
\hbar\omega_{31}\Phi_3+V_3(\mathbf{r})\Phi_3-\hbar\Omega_c\Phi_2-\hbar
 g\mathcal{E}\Phi_1,
\label{eq:phi3}
\\\hat{K}\Phi_2 & = &
\hbar\omega_{21}\Phi_2+V_2(\mathbf{r})\Phi_2-\hbar\Omega_c^{\ast}\Phi_3\:,
\label{eq:phi2}
\end{eqnarray}
with
\begin{equation}
\hat{K}=i\hbar\frac{\partial}{\partial t}+\frac{\hbar^2}{2m}\nabla^2\:.
\end{equation}
Here $m$ is the atomic mass, $V_j(\mathbf{r})$ is the trapping potential for an
atom in the electronic state $j$ ($j=1,2,3$), $\Omega_c$ is the Rabi-frequency
of the control laser driving the transition $|2\rangle\rightarrow|3\rangle$;
$\omega_{21}=\omega_2-\omega_1+\omega_c-\omega$ and
$\omega_{31}=\omega_3-\omega_1-\omega$ are, respectively, the frequencies of
electronic detuning from the two- and one-photon resonances. In the limit of
infinite mass we can ignore the spatial derivatives in these equations which
then reduce to the familiar equations for the probability amplitudes describing
the coherent excitation of three-level atoms in the $\Lambda$ configuration
\cite{scully97:_quant_optic}. For us, however, retention of this center of mass
motion is essential in order to describe light dragging effects.  Note also
that the equation of motion (\ref{eq:phi1}) for $\Phi_1$ does not explicitly
accommodate collisions between the ground-state atoms. In the case of a BEC
where the state $1$ is mostly populated, the collisional effects can be 
included replacing $V_1(\mathbf{r})$ by
$V_1(\mathbf{r})+g_{11}|\Phi_1|^2$ in Eq.~(\ref{eq:phi1}) to yield a
mean-field equation for the condensate wave-function $\Phi_1$, where 
$g_{11}=4\pi\hbar^{2}a_{11}/m$ and $a_{11}$ is the scattering length between 
the condensate atoms in state $1$. In
Eqs.~(\ref{eq:phi1})-(\ref{eq:phi2}) the coupling with the probe and control
fields has been written using the rotating wave approximation. Therefore the
last term in Eq.~(\ref{eq:phi1}) has a positive frequency part of the probe
field (i.e. $\mathcal{E}^{*}$), whereas the last term in Eq.~(\ref{eq:phi3})
has a negative frequency part ($\mathcal{E}$). For the same reason,
Eq.~(\ref{eq:phi3}) contains a factor $\Omega_c$, whereas Eq.~(\ref{eq:phi2})
contains a factor $\Omega_c^{\ast}$.

Equation~(\ref{eq:phi2}) relates $\Phi_3$ to $\Phi_2$ as:
\begin{equation}
\Phi_3=\frac{1}{\hbar\Omega_c^*}\left(-\hat{K}+V_2(\mathbf{r})+\hbar\omega_{
21}\right)\Phi_2\:.
\label{eq:phi3-phi2}
\end{equation}
Initially the atoms are in the ground level $1$ and the Rabi frequency of the
probe field is considered to be much smaller than $\Omega_c$. Consequently one
can neglect the last term in Eq.~(\ref{eq:phi1}) that causes depletion of the
ground level $1$, giving $\hat{K}\Phi_1=V_1(\mathbf{r})\Phi_1$. In the case of
a BEC the wave-function $\Phi_1=\sqrt{n}\exp(iS_1)$ represents an incident
variable determining the atomic density $n$ and velocity field $\hbar \nabla
S_{1}/m$.

\section{Propagation of probe beam}

\subsection{Adiabatic approximation}

Suppose the control and probe beams are close to the two-photon resonance.
Application of the two beams cause electromagnetically induced transparency
(EIT)
\cite{harris90,boller91,arimondo96:_progr_optic,harris97,scully97:_quant_optic,bergmann98,hau99,kash99,budker99}
in which the transitions $|2\rangle\rightarrow|3\rangle$ and
$|1\rangle\rightarrow|3\rangle$ interfere destructively preventing population of
the excited state $3$. The adiabatic approximation is obtained neglecting the
excited state population in Eq.~(\ref{eq:phi3}) to yield:
\begin{equation}
\Phi_2=-g\frac{\Phi_1}{\Omega_c}\mathcal{E}\,.
\label{eq:phi2-adiabat}
\end{equation}
Equations (\ref{eq:E-equation}), (\ref{eq:phi3-phi2}) and
(\ref{eq:phi2-adiabat}) provide a closed set of the equations for the electric
field amplitude $\mathcal{E}$:
\begin{eqnarray}
&&\left(\frac{\partial}{\partial
t}-i\frac{c^2}{2\omega}\nabla^2-i\frac{\omega}{2}\right)\mathcal{E}=\nonumber \\ &&-i\frac{
g^2\Phi_1^*}{\hbar\Omega_c^*}\left(-\hat{K}+V_2(\mathbf{r})+\hbar\omega_{
21}\right)\frac{\Phi_1\mathcal{E}}{\Omega_c}\:.
\label{eq:E-2}
\end{eqnarray}
This equation suffices to describe a wide variety of phenomena.  In particular
it can be used to model light storage by introducing time-dependence in
$|\Omega_c|$ \cite{fleischhauer00} and light dragging due to spatial variation
of $\Omega_c$ and/or $\Phi_1$.  Here we shall be especially interested in image
rotation associated with spatial variations of $\Phi_1$, corresponding to a
moving medium and with light beams carrying orbital angular momentum so that
$\mathcal{E} \propto e^{i\ell\phi}$.

\subsection{Co-propagating control and probe beams}

Suppose the probe and control beams co-propagate:
\begin{eqnarray}
\mathcal{E}(\mathbf{r},t) & = &
\tilde{\mathcal{E}}(\mathbf{r},t)e^{ikz}\:,
\label{eq:E-copoprogat}
\\
\Omega_c(\mathbf{r},t) & = &
\Omega(\mathbf{r},t)e^{ik_cz}\:,
\label{eq:Omega-copropagat}
\end{eqnarray}
where $k_c=\omega_c/c$ is the wave-number of the control beam. For paraxial
beams the amplitudes $\tilde{\mathcal{E}}(\mathbf{r},t)$ and
$\Omega(\mathbf{r},t)$ depend only weakly on the propagation direction $z$.
Furthermore the amplitude of the control field
$\Omega(\mathbf{r},t)$ is considered to change weakly in the
transverse direction as well. Neglecting the spatial dependence of $\Omega(\mathbf{r},t)$,
the equation (\ref{eq:E-2}) for the probe field takes the form
\begin{equation}
\frac{\partial}{\partial
t}\tilde{\mathcal{E}}+v_g\left[\frac{\partial}{\partial
z}+\left(\frac{1}{v_g}-\frac{1}{c}\right)(\mathbf{v}_a\cdot\nabla+i\delta)
-i\frac{1}{2k}\nabla_{\bot}^2\right]\tilde{\mathcal{E}}=0\,,
\label{eq:E-3}
\end{equation}
where we have replaced $\nabla^2$ by its transverse part
$\nabla_{\bot}^2=\partial^2/\partial x^2+\partial^2/\partial y^2$ because of
the paraxial approximation.  Here
\begin{equation}
v_g=\frac{c|\Omega_c(\mathbf{r},t)|^2}{|\Omega_c(\mathbf{r},t)|^2+g^2n(\mathbf{
r})}
\label{eq:v-g}
\end{equation}
is the radiative group velocity,
\begin{equation}
\mathbf{v}_a=-\frac{i\hbar}{m}\frac{1}{\Phi_1}\nabla\Phi_1=\frac{\hbar}{
m}\nabla S_1-\frac{i\hbar}{2m}\frac{1}{n}\nabla n
\label{eq:v}
\end{equation}
is a complex atomic velocity field, and
\begin{equation}
\delta=\omega_{21}+\frac{1}{\hbar}(V_2(\mathbf{r})-V_1(\mathbf{r}))
\label{eq:delta}
\end{equation}
is the two-photon frequency mismatch. As the beams co-propagate, the ordinary
two-photon Doppler shift $\mathbf{v}_a\cdot\hat{\mathbf{z}}(k-k_c)$ and two
photon recoil $\hbar(k-k_c)^2/2m$ are of little importance and hence have been
omitted.  Furthermore, the radiative velocity $v_g$ is much larger than the
velocity of the atomic recoil $v_{\mathrm{rec}}\equiv\hbar k/m$ , so the latter
$v_{\mathrm{rec}}$ has been neglected in the equation of motion (\ref{eq:E-3}).

The term with the spatial derivative $\partial/\partial z$ in
Eq.~(\ref{eq:E-3}) describes the radiative propagation along the $z$-axis with
the group velocity $v_g$. On the other hand, the term containing
$\mathbf{v}_a\cdot\nabla$ represents the dragging of light by a moving medium.
The amount of such a dragging is determined by the difference in the group
velocity and the speed of light. For fast light there is no dragging, but for
slow light a substantial dragging is possible.

\subsection{Rotating medium}

Consider a probe beam carrying an orbital angular moment (OAM) $\hbar \ell$ per
photon along the propagation axis $\tilde{\mathcal{E}}(\mathbf{r},t)\sim
e^{i\ell \varphi}$, where $\varphi$ is the azimuthal angle.  Suppose the medium
is rotating as a rigid body with a frequency $\omega_{\mathrm{rot}}$ in the
$xy$ plane. For instance, an atomic vortex lattice exhibits such rigid body rotation on a
coarse grained scale \cite{pitaevskii03}. In such a situation the linear
velocity is $\mathbf{v}_a=\omega_{\mathrm{rot}}\rho \mathbf{e}_{\varphi}$,
where $\mathbf{e}_{\varphi}$ is the unit vector along the azimuthal direction
and $\rho=\sqrt{x^2+y^2}$ is the cylindrical radius. Under these conditions the
dragging term featured in the equation of motion (\ref{eq:E-3}) takes the form
\begin{equation}
\mathbf{v}_a\cdot\nabla\tilde{\mathcal{E}}(\mathbf{r},t)=il\omega_{\mathrm{
rot}}\tilde{\mathcal{E}}(\mathbf{r},t)\,.
\label{eq:v-a-rotating}
\end{equation}
Suppose the atomic density (and hence the group velocity) is approximately
constant in the transverse direction \cite{comment}, so that the group velocity depends on the propagation coordinate only: $v_g= v_{g}(z)$. Using Eqs.~(\ref{eq:E-3}) and
(\ref{eq:v-a-rotating}), the total phase shift then acquired by the beam
propagating along the $z$-axis is
\begin{equation}
\theta_\ell=(\ell\omega_{\mathrm{rot}}+\delta)\int_{z_0}^z\left(\frac{1}{v_g(z^{
\prime})}-\frac{1}{c}\right)dz^{\prime}\,,
\label{eq:Delta-theta}
\end{equation}
where $z_0$ is the entrance point into the medium. The first term in
$(\ell\omega_{\mathrm{rot}}+\delta)$ represents the rotational frequency shift
due to the orbital angular momenta $\hbar \ell$ of the probe photons. If the control and
probe fields are circularly polarised, the second term accommodates the
rotational two-photon frequency shift $\pm2\omega_{\mathrm{rot}}$, as well as
the residual detuning $\delta^{\prime}$
\begin{equation}
\delta=\pm2\omega_{\mathrm{rot}}+\delta^{\prime}\,,
\end{equation}
where the sign of $\pm2\omega_{\mathrm{rot}}$ depends on the polarisations of
the control and probe field. The factor $2$ appears because the control and
probe beams are considered to have opposite circular polarisations leading a
double frequency shift. In the slow-light experiments involving a $\Lambda$
scheme \cite{liu01,phillips01} it is quite common for the control and probe
fields to have the opposite circular polarisations which drive the atomic
transitions with $\Delta m=\pm1$. The residual detuning $\delta^{\prime}$
depends neither on the orbital angular momentum of a photon nor on its
polarisation and is hence of little importance for the subsequent discussion.

\subsection{Polarisation rotation of slow light with OAM}

A characteristic feature of an EIT medium is that only one circularly polarised
component of light (say left polarised light) is slowed down. The right
circularly polarised light has a forbidden resonant transition to the excited
state and is propagating with a group velocity close to the speed of light in
vacuum. If the medium rotates and the probe beam carries an  OAM, the left 
circularly polarised beam of light acquires a large
rotational frequency shift because of its small group velocity. On the other
hand, the right polarised light is a fast light and hence experiences almost no
rotational frequency shift. If the incoming light beam is linearly polarised,
this leads to the rotation of the polarisation plane by an angle
$\alpha_\ell=(\theta_\ell+\theta_{\mathrm{fast}})/2$, where
$\theta_{\mathrm{fast}}$ is the dynamical phase shift acquired by the fast
(right polarised) light, $\theta_\ell$ being the phase shift acquired by the
slow (left polarised) light given by Eq.~(\ref{eq:Delta-theta}). To extract a
pure rotational phase shift, we suggest measuring the difference in the angles
of the polarisation plane rotation for two consecutive probe beams
characterised by opposite OAM
\begin{equation}
\Delta\alpha_\ell=\alpha_\ell-\alpha_{-\ell}=\ell\omega_{\mathrm{rot}}\int_{
z_0}^z\left(\frac{1}{v_g(z^{\prime})}-\frac{1}{c}\right)dz^{\prime}\,,
\label{eq:Delta-alpha}
\end{equation}
where we have taken advantage of the fact that $\theta_{\mathrm{fast}}$
and $\delta$ do not depend on the winding number $\ell$.

The angle $\Delta\alpha_l$ is proportional to the rotation frequency of the
medium $\omega_{\mathrm{rot}}$ and to the winding number $\ell$. Therefore the
effect is due to the orbital angular momentum of slow light.  Furthermore, the
angle $\Delta\alpha_\ell$  is inversely proportional to the group velocity of
light. This suggests that for slow light the angle $\Delta\alpha_l$ should be
large enough to be detectable.  Let us make some numerical estimates for
typical BEC / slow light experimental parameters. In the experiment by Hau
\textit{et al} \cite{hau99} the radiative group velocity is
$v_g=17\,\mathrm{m/s}$, whereas the length of a cigar shaped BEC is $\Delta
z=z_1-z_0=0.3\,\mathrm{mm}=3\times10^{-4}\,\mathrm{m}$. Typically the
transverse frequency of an elongated trap is $\omega\sim2\pi\times10^2$, so the
rotation frequency $\omega_{\mathrm{rot}}$ should be of the same order. Thus
the difference in the angle of the polarisation plane rotation is
$\Delta\alpha_\ell\approx4\pi \ell \times10^{-5}$. This is a small angle, yet by no
means beyond current experimental expertise \cite{Lintz06}.  For fast light the
corresponding angle $\Delta\alpha_\ell$ should be far too small to be observed.

\section{Conclusions}

In the present paper we have considered the propagation of a slow light with an
orbital angular momentum in a moving atomic medium. We have derived a general
equation of motion and applied it to propagation of a slow light with an
orbital angular momentum in a rotating medium. We have shown that the orbital
angular momentum of slow light manifest itself in a rotation of the
polarisation plane of a linear polarised light. To extract a pure rotational
phase shift, we suggest to measure the difference in the angles of the
polarisation plane rotation for two consecutive beams of probe light
characterised by opposite OAM. The differential angle $\Delta\alpha_\ell$ is
proportional to the rotation frequency of the medium $\omega_{\mathrm{rot}}$,
as well as to the winding number $\ell$. The effect can be used as a tool for
measuring the rotation frequency $\omega_{\mathrm{rot}}$ of the medium and/or
the orbital angular momentum of slow light.

\begin{acknowledgments}
This work was supported by the Royal Society, the UK Engineering and Physical
Sciences Research Council, the Royal Society of Edinburgh and by the Wolfson
Foundation.
\end{acknowledgments}


\begin{thebibliography}{31}
\bibitem[{com()}]{comment}{In many BEC experiments the external trap
is harmonic.  With increasing rotation the trapped atomic cloud acquires a
pancake shape (with holes if the vortex lattice is formed). Therefore the
density is on average constant over a reasonably large area in the central
region of the trap.}
\end{thebibliography}

\end{document}